

\font\bigbf=cmbx10 scaled\magstep4

\font\fourteenbf=cmbx10 scaled\magstep2

\font\twelvesy=cmsy10 scaled 1200
\font\twelveex=cmex10 scaled 1200
\font\twelvesc=cmcsc10 scaled 1200

\font\twelverm=cmr12
\font\twelvei=cmmi12
\font\twelvebf=cmbx12
\font\twelvesl=cmsl12
\font\twelvett=cmtt12
\font\twelveit=cmti12
\font\twelvesf=cmss12
\font\tensc=cmcsc10
\skewchar\twelvei='177
\skewchar\twelvesy='60
 \def\twelvepoint{\normalbaselineskip=12.4pt plus 0.1pt minus 0.1pt
  \abovedisplayskip 12.4pt plus 3pt minus 9pt
  \belowdisplayskip 12.4pt plus 3pt minus 9pt
  \abovedisplayshortskip 0pt plus 3pt
  \belowdisplayshortskip 7.2pt plus 3pt minus 4pt
  \smallskipamount=3.6pt plus1.2pt minus1.2pt
  \medskipamount=7.2pt plus2.4pt minus2.4pt
  \bigskipamount=14.4pt plus4.8pt minus4.8pt
  \def\rm{\fam0\twelverm}
  \def\it{\fam\itfam\twelveit}
  \def\sl{\fam\slfam\twelvesl}
  \def\bf{\fam\bffam\twelvebf}
  \def\mit{\fam 1}
  \def\cal{\fam 2}
  \def\sc{\twelvesc}
  \def\tt{\twelvett}
  \def\sf{\twelvesf}
  \textfont0=\twelverm   \scriptfont0=\tenrm   \scriptscriptfont0=\sevenrm
  \textfont1=\twelvei    \scriptfont1=\teni    \scriptscriptfont1=\seveni
  \textfont2=\twelvesy   \scriptfont2=\tensy   \scriptscriptfont2=\sevensy
  \textfont3=\twelveex   \scriptfont3=\twelveex  \scriptscriptfont3=\twelveex
  \textfont\itfam=\twelveit  \textfont\slfam=\twelvesl
  \textfont\bffam=\twelvebf  \scriptfont\bffam=\tenbf
  \scriptscriptfont\bffam=\sevenbf
  \normalbaselines\rm}

\let\medtype=\tenpoint

\def\beginlinemode{\endmode
  \begingroup\parskip=0pt \obeylines\def\\{\par}\def\endmode{\par\endgroup}}
\def\beginparmode{\endmode
  \begingroup \def\endmode{\par\endgroup}}
\let\endmode=\par
{\obeylines\gdef\
{}}
\def\singlespace{\baselineskip=\normalbaselineskip}

\def\oneandahalfspace{\baselineskip=\normalbaselineskip
  \multiply\baselineskip by 3 \divide\baselineskip by 2}
\def\doublespace{\baselineskip=\normalbaselineskip \multiply\baselineskip by 2}

\pageno=0
\newcount\firstpageno\firstpageno=2
\footline={\ifnum\pageno<\firstpageno{\hfil}%
\else{\hfil\twelverm\folio\hfil}\fi}
\def\toppageno{\global\footline={\hfil}\global\headline
  ={\ifnum\pageno<\firstpageno{\hfil}\else{\hfil\twelverm\folio\hfil}\fi}}
\let\rawfootnote=\footnote          
\def\footnote#1#2{{\rm\singlespace\parindent=0pt\parskip=0pt
  \rawfootnote{#1}{#2\hfill\vrule height 0pt depth 6pt width 0pt}}}
\def\raggedcenter{\leftskip=3em plus 12em \rightskip=\leftskip
  \parindent=0pt \parfillskip=0pt \spaceskip=.3333em \xspaceskip=.5em
  \pretolerance=9999 \tolerance=9999
  \hyphenpenalty=9999 \exhyphenpenalty=9999 }
\def\dateline{\rightline{\ifcase\month\or
  January\or February\or March\or April\or May\or June\or
  July\or August\or September\or October\or November\or December\fi
  \space\number\year}}
\def\ldateline{\leftline{\ifcase\month\or
  January\or February\or March\or April\or May\or June\or
  July\or August\or September\or October\or November\or December\fi
  \space\number\year}}

\hsize=16.5truecm
\vsize=23.0truecm
\hoffset=0truein
\voffset=0truein
\parskip=\medskipamount
\def\\{\cr}
\twelvepoint \oneandahalfspace
\overfullrule=0pt     

\newcount\timehour
\newcount\timeminute
\newcount\timehourminute
\def\daytime{\timehour=\time\divide\timehour by 60
  \timehourminute=\timehour\multiply\timehourminute by -60
  \timeminute=\time\advance\timeminute by \timehourminute
  \number\timehour:\ifnum\timeminute<10{0}\fi\number\timeminute}
\def\today{\number\day\space\ifcase\month\or Jan\or Feb\or Mar
  \or Apr\or May\or Jun\or Jul\or Aug\or Sep\or Oct\or
  Nov\or Dec\fi\space\number\year}

\def\title{\null\vskip 5pt plus 0.2fill
   \beginlinemode \doublespace \raggedcenter \bigbf}
\def\author{\vskip 3pt plus 0.2fill \beginlinemode  \doublespace
   \raggedcenter \fourteenbf}
\def\affil{\vskip 3pt plus 0.1fill
   \beginlinemode \oneandahalfspace \raggedcenter \it}

\def\abstract{\vskip 3pt plus 0.3fill {\raggedcenter{\rm ABSTRACT}}
   \beginparmode \narrower \oneandahalfspace }
\def\body{\beginparmode}
\def\endtopmatter{\endpage\body}

\def\head#1{ \vfill\eject\vskip 0.4truein  {\immediate\write16{#1}
   \raggedcenter {\fourteenbf #1} \par }
   \nobreak\vskip 0truein\nobreak}
\def\subhead#1{\vskip 0.25truein
  {\leftline {\bf #1} \par}
   \nobreak\vskip 0truein\nobreak}
\def\subsubhead#1{ \vskip 0.25truein  {\leftline {\bf #1} \par}
   \nobreak\vskip 0truein\nobreak}

\newcount\figno\figno=0
\newcount\chapno\chapno=0
\newcount\subchapno\subchapno=0
\newcount\subsubchapno\subsubchapno=0

\outer\def\chap#1{\global\advance\chapno by 1
\global\figno=0
\global\subchapno=0 \global\subsubchapno=0
 \head{\the \chapno. #1} \taghead{\the\chapno.} }

\outer\def\subchap#1{\global\advance\subchapno by 1
 \global\subsubchapno=0
 \subhead{\the\chapno.\the\subchapno. #1}  }

\outer\def\subsubchap#1{\global\advance\subsubchapno by 1
 \subsubhead{\the\chapno.\the\subchapno.\the\subsubchapno. #1} }

\outer\def\Fig #1 #2 #3 #4 #5{\global\advance\figno by 1
\midinsert \vglue#1cm \hskip#2cm
\special{picture #3 scaled #4} \smallskip
\medtype \singlespace \parshape=2 2.25cm 12cm 3.75cm 10.5cm
{ {\bf Figure \the\chapno.\the\figno.}  \rm #5.}  \endinsert}

\outer\def\Fign #1 #2 #3 #4 {\global\advance\figno by 1
\midinsert  \vglue#1cm \hskip#2cm
\special{picture #3 scaled #4} \smallskip
{\centerline{\tenbf Figure \the\chapno.\the\figno. }}
\endinsert}

\def\beneathrel#1\under#2{\mathrel{\mathop{#2}\limits_{#1}}}

\def\references  {\head{References}
   \beginparmode\singlespace    
   \frenchspacing \parindent=0pt
      \parskip=\smallskipamount 
 \everypar{\hangindent=20pt  \hangafter=1}}

\def\refstylesprin{
  \gdef\refto##1{$\rm [##1]$}
  \gdef\refis##1{{[##1]\ }}
 \gdef\journal##1,##2,##3,##4&{{\sl ##1 }{\bf ##2 }(##4) ##3}}

\refstylesprin

\def\cmp{\journal Comm. Math. Phys.}

\def\np{\journal Nucl. Phys.}
\def\pl{\journal Phys. Lett.}

\def\endreferences{\body}
\def\ref#1{Ref.~#1}               
\def\Ref#1{Ref.~[#1]}               
\def\[#1]{[\cite{#1}]}
\def\cite#1{{#1}}
\def\(#1){(\call{#1})}
\def\call#1{{#1}}
\def\taghead#1{}
\def\frac#1#2{{#1 \over #2}}
\def\13{{\frac13}}
\def\14{{\frac14}}
\def\12{{1\over2}}

\def\ie{{\it i.e.\ }}
\def\etal{{\it et al.}}

\def\sla{\raise.15ex\hbox{$/$}\kern -.57em}
\def\leaderfill{\leaders\hbox to 1em{\hss.\hss}\hfill}
\def\twiddle{\lower.9ex\rlap{$\kern -.1em\scriptstyle\sim$}}
\def\bigtwiddle{\lower1.ex\rlap{$\sim$}}
\def\gtwid{\mathrel{\raise.3ex\hbox{$>$\kern -.75em\lower1ex\hbox{$\sim$}}}}
\def\ltwid{\mathrel{\raise.3ex\hbox{$<$\kern -.75em\lower1ex\hbox{$\sim$}}}}
\def\square{\kern1pt\vbox{\hrule height 1.2pt\hbox{\vrule width 1.2pt\hskip 3pt
   \vbox{\vskip 6pt}\hskip 3pt\vrule width 0.6pt}\hrule height 0.6pt}\kern1pt}
\def\tdot#1{\mathord{\mathop{#1}\limits^{\kern2pt\ldots}}}

\def\pmb#1{\setbox0=\hbox{#1}%
  \kern -.025em\copy0\kern -\wd0
  \kern  .05em\copy0\kern -\wd0
  \kern -.025em\raise.0433em\box0 }


\def\ket#1{\left| #1\right\rangle}

\def\@versim#1#2{\lower0.2ex\vbox{\baselineskip\z@skip\lineskip\z@skip
  \lineskiplimit\z@\ialign{$\m@th#1\hfil##\hfil$\crcr#2\crcr\sim\crcr}}}
\def\gsim{\mathrel{\mathpalette\@versim>}}
\def\lsim{\mathrel{\mathpalette\@versim<}}

\def\rchi{\raise2pt\hbox{$\chi$}}

\def\Re{{\cal R \mskip-4mu \lower.1ex \hbox{\it e}}}
\def\Im{{\cal I \mskip-5mu \lower.1ex \hbox{\it m}}}

\def\rlh{\scriptstyle{\rightharpoonup\hskip-8pt{\leftharpoondown}}}
\def\harpar{\partial\hskip-8pt\raise9pt\hbox{$\rlh$}}

\def\bar{\overline}

\def\endpage {\vfill\eject}
\def\endpaper{\endmode\vfill\supereject}
\def\endit {\endpaper\end}

\catcode`@=11
\newcount\tagnumber\tagnumber=0
\immediate\newwrite\eqnfile
\newif\if@qnfile\@qnfilefalse
\def\write@qn#1{}
\def\writenew@qn#1{}
\def\w@rnwrite#1{\write@qn{#1}\message{#1}}
\def\@rrwrite#1{\write@qn{#1}\errmessage{#1}}

\def\taghead#1{\gdef\t@ghead{#1}\global\tagnumber=0}
\def\t@ghead{}

\expandafter\def\csname @qnnum -3\endcsname
  {{\t@ghead\advance\tagnumber by -3\relax\number\tagnumber}}
\expandafter\def\csname @qnnum -2\endcsname
  {{\t@ghead\advance\tagnumber by -2\relax\number\tagnumber}}
\expandafter\def\csname @qnnum -1\endcsname
  {{\t@ghead\advance\tagnumber by -1\relax\number\tagnumber}}
\expandafter\def\csname @qnnum0\endcsname
  {\t@ghead\number\tagnumber}
\expandafter\def\csname @qnnum+1\endcsname
  {{\t@ghead\advance\tagnumber by 1\relax\number\tagnumber}}
\expandafter\def\csname @qnnum+2\endcsname
  {{\t@ghead\advance\tagnumber by 2\relax\number\tagnumber}}
\expandafter\def\csname @qnnum+3\endcsname
  {{\t@ghead\advance\tagnumber by 3\relax\number\tagnumber}}

\def\equationfile{%
  \@qnfiletrue\immediate\openout\eqnfile=\jobname.eqn%
  \def\write@qn##1{\if@qnfile\immediate\write\eqnfile{##1}\fi}
  \def\writenew@qn##1{\if@qnfile\immediate\write\eqnfile
     {\noexpand\tag{##1} = (\t@ghead\number\tagnumber)}\fi}
     }
\def\callall#1{\xdef#1##1{#1{\noexpand\call{##1}}}}
\def\call#1{\each@rg\callr@nge{#1}}

\def\each@rg#1#2{{\let\thecsname=#1\expandafter\first@rg#2,\end,}}
\def\first@rg#1,{\thecsname{#1}\apply@rg}
\def\apply@rg#1,{\ifx\end#1\let\next=\relax%
\else,\thecsname{#1}\let\next=\apply@rg\fi\next}

\def\callr@nge#1{\calldor@nge#1-\end -}
\def\callr@ngeat#1\end -{#1}
\def\calldor@nge#1-#2-{\ifx\end#2\@qneatspace#1 %
  \else\calll@@p{#1}{#2}\callr@ngeat\fi}
\def\calll@@p#1#2{\ifnum#1>#2{\@rrwrite{Equation range #1-#2\space is bad.}
\errhelp{If you call a series of equations by the notation M-N, then M and
N must be integers, and N must be greater than or equal to M.}}\else%
{\count0=#1\count1=#2\advance\count1 by1\relax\expandafter\@qncall\the\count0,%
  \loop\advance\count0 by1\relax%
    \ifnum\count0<\count1,\expandafter\@qncall\the\count0,%
  \repeat}\fi}

\def\@qneatspace#1#2 {\@qncall#1#2,}
\def\@qncall#1,{\ifunc@lled{#1}{\def\next{#1}\ifx\next\empty\else
  \w@rnwrite{Equation number \noexpand\(>>#1<<) has not been defined yet.}
  >>#1<<\fi}\else\csname @qnnum#1\endcsname\fi}

\let\eqnono=\eqno
\def\eqno(#1){\tag#1}
\def\tag#1$${\eqnono(\displayt@g#1 )$$}

\def\aligntag#1\endaligntag
  $${\gdef\tag##1\\{&(##1 )\cr}\eqalignno{#1\\}$$
  \gdef\tag##1$${\eqnono(\displayt@g##1 )$$}}

\def\eqalignno#1{\displ@y \tabskip\centering
  \halign to\displaywidth{\hfil$\displaystyle{##}$\tabskip\z@skip
    &$\displaystyle{{}##}$\hfil\tabskip\centering
    &\llap{$\displayt@gpar##$}\tabskip\z@skip\crcr
    #1\crcr}}

\def\displayt@gpar(#1){(\displayt@g#1 )}
\def\displayt@g#1 {\rm\ifunc@lled{#1}\global\advance\tagnumber by1
        {\def\next{#1}\ifx\next\empty\else\expandafter
        \xdef\csname @qnnum#1\endcsname{\t@ghead\number\tagnumber}\fi}%
  \writenew@qn{#1}\t@ghead\number\tagnumber\else
        {\edef\next{\t@ghead\number\tagnumber}%
        \expandafter\ifx\csname @qnnum#1\endcsname\next\else
        \w@rnwrite{Equation \noexpand\tag{#1} is a duplicate number.}\fi}%
  \csname @qnnum#1\endcsname\fi}

\def\ifunc@lled#1{\expandafter\ifx\csname @qnnum#1\endcsname\relax}
\let\@qnend=\end\gdef\end{\if@qnfile
\immediate\write16{Equation numbers written on []\jobname.EQN.}\fi\@qnend}
\catcode`@=12

\catcode`@=11
\newcount\r@fcount \r@fcount=0
\newcount\r@fcurr
\immediate\newwrite\reffile
\newif\ifr@ffile\r@ffilefalse
\def\w@rnwrite#1{\ifr@ffile\immediate\write\reffile{#1}\fi\message{#1}}

\def\writer@f#1>>{}
\def\referencefile{
  \r@ffiletrue\immediate\openout\reffile=\jobname.ref%
  \def\writer@f##1>>{\ifr@ffile\immediate\write\reffile%
    {\noexpand\refis{##1} = \csname r@fnum##1\endcsname = %
     \expandafter\expandafter\expandafter\strip@t\expandafter%
     \meaning\csname r@ftext\csname r@fnum##1\endcsname\endcsname}\fi}%
  \def\strip@t##1>>{}}

\def\citeall#1{\xdef#1##1{#1{\noexpand\cite{##1}}}}
\def\cite#1{\each@rg\citer@nge{#1}}     

\def\each@rg#1#2{{\let\thecsname=#1\expandafter\first@rg#2,\end,}}
\def\first@rg#1,{\thecsname{#1}\apply@rg}     
\def\apply@rg#1,{\ifx\end#1\let\next=\relax
\else,\thecsname{#1}\let\next=\apply@rg\fi\next}

\def\citer@nge#1{\citedor@nge#1-\end -}     
\def\citer@ngeat#1\end -{#1}
\def\citedor@nge#1-#2-{\ifx\end#2\r@featspace#1 
  \else\citel@@p{#1}{#2}\citer@ngeat\fi}     
\def\citel@@p#1#2{\ifnum#1>#2{\errmessage{Reference range #1-#2\space is bad.}
    \errhelp{If you cite a series of references by the notation M-N, then M and
    N must be integers, and N must be greater than or equal to M.}}\else%
 {\count0=#1\count1=#2\advance\count1
by1\relax\expandafter\r@fcite\the\count0,%
  \loop\advance\count0 by1\relax
    \ifnum\count0<\count1,\expandafter\r@fcite\the\count0,%
  \repeat}\fi}

\def\r@featspace#1#2 {\r@fcite#1#2,}     
\def\r@fcite#1,{\ifuncit@d{#1}          
    \expandafter\gdef\csname r@ftext\number\r@fcount\endcsname%
    {\message{Reference #1 to be supplied.}\writer@f#1>>#1 to be supplied.\par
     }\fi%
  \csname r@fnum#1\endcsname}

\def\ifuncit@d#1{\expandafter\ifx\csname r@fnum#1\endcsname\relax%
\global\advance\r@fcount by1%
\expandafter\xdef\csname r@fnum#1\endcsname{\number\r@fcount}}

\let\r@fis=\refis               
\def\refis#1#2#3\par{\ifuncit@d{#1}
    \w@rnwrite{Reference #1=\number\r@fcount\space is not cited up to now.}\fi%
  \expandafter\gdef\csname r@ftext\csname r@fnum#1\endcsname\endcsname%
  {\writer@f#1>>#2#3\par}}

\def\r@ferr{\endreferences\errmessage{I was expecting to see
\noexpand\endreferences before now;  I have inserted it here.}}
\let\r@ferences=\references
\def\references{\r@ferences\def\endmode{\r@ferr\par\endgroup}}

\let\endr@ferences=\endreferences
\def\endreferences{\r@fcurr=0
  {\loop\ifnum\r@fcurr<\r@fcount
    \advance\r@fcurr by 1\relax\expandafter\r@fis\expandafter{\number\r@fcurr}%
    \csname r@ftext\number\r@fcurr\endcsname%
  \repeat}\gdef\r@ferr{}\endr@ferences}

\let\r@fend=\endpaper\gdef\endpaper{\ifr@ffile
\immediate\write16{Cross References written on []\jobname.REF.}\fi\r@fend}
\catcode`@=12
\citeall\refto \citeall\ref \citeall\Ref

\catcode`@=11
\newwrite\tocfile\openout\tocfile=\jobname.toc
\newlinechar=`^^J
\write\tocfile{\string\input\space ft700^^J
  \string\pageno=-1\string\firstpageno=-1000\string\singlespace
  \string\null\string\vfill\string\centerline{TABLE OF CONTENTS}^^J
  \string\vskip 0.5 truein\string\rightline{\string\underbar{Page}}\smallskip}

\def\tocitem#1{
  \t@cskip{#1}\bigskip}
\def\tocitemitem#1{
  \t@cskip{\quad#1}\medskip}
\def\tocitemitemitem#1{
  \t@cskip{\qquad#1}\smallskip}
\def\tocitemall#1{
  \xdef#1##1{#1{##1}\noexpand\tocitem{##1}}}
\def\tocitemitemall#1{
  \xdef#1##1{#1{##1}\noexpand\tocitemitem{##1}}}
\def\tocitemitemitemall#1{
  \xdef#1##1{#1{##1}\noexpand\tocitemitemitem{##1}}}

\def\t@cskip#1#2{
  \write\tocfile{\string#2\string\line{^^J
  #1\string\leaderfill\space\number\folio}}}

%

\def\t@cproduce{
  \write\tocfile{\string\vfill\string\vfill\string\supereject\string\end}
  \closeout\tocfile
  \immediate\write16{Table of Contents written on []\jobname.TOC.}}

\let\t@cend=\endpaper\def\endpaper{\t@cproduce\t@cend}
\catcode`@=12
\tocitemall\head          
\tocitemitemall\subhead
\tocitemitemitemall\subsubhead

\rightline{ \rm{UGVA--DPT~1991/08--739} }
\rightline{ \rm{CERN--TH.6200/91}}

\title{A New Solution to the Star--Triangle Equation Based on U$_q$(sl(2)) at
Roots of Unit}

\vskip.5cm

\author{C\'esar G\'omez$^{*\dagger}$}

\affil{D\'ept. Physique Th\'eorique, Universit\'e de Gen\`eve, %
CH--1211  Gen\`eve 4
\footnote{}{$^*$Supported in part by the  Fonds National Suisse pour %
 la Recherche Scientifique.}}

 \author{ Germ\'an Sierra$^\dagger$}
\affil{Theory Division, CERN, CH--1211 Gen\`eve 23}
\footnote{}{$^\dagger$Permanent address: Instituto de F\'\i sica  %
Fundamental, CSIC, Serrano 123,  Madrid 28006.}
\vskip1cm
\centerline{\bf ABSTRACT}

\beginparmode\narrower{ We find new solutions to the Yang--Baxter equation in
terms of the intertwiner matrix for semi-cyclic representations of the quantum
group  $U_q(s\ell(2))$ with $q= e^{2\pi i/N}$.  These intertwiners serve to
define the Boltzmann weights of a lattice model, which shares some similarities
with the chiral Potts model. An alternative interpretation of these Boltzmann
weights is as scattering matrices of solitonic structures whose kinematics is
entirely governed by the quantum group. Finally, we consider the limit
$N\to\infty$ where we find an infinite--dimensional representation of the braid
group, which may give rise to an invariant of knots and links.}

\endtopmatter
\taghead{{}}\global\chapno=1

\subhead{1. Introduction}
The $N$--chiral  Potts model \refto{1,2} is a solvable lattice model
satisfying the star--triangle relation. Their Boltzmann weights are meromorphic
functions on an algebraic curve of genus $N^3-2N^2+1$. These models are the
first example of solutions to the Yang--Baxter equation with the spectral
parameters living on a curve of genus greater than one. Recent developments
\refto{3,4} strongly indicate that quantum groups at roots of unit \refto{5}
characterize the underlying symmetry of these models. More precisely, it was
shown in reference \refto{4} that the intertwiner $R$ matrix for cyclic
representations of the affine Hopf algebra $U_q(\hat{s\ell} (2))$ at $q$ an
$N$-th root of unit, admit a complete factorization in terms of the Boltzmann
weights of the $N$--chiral Potts model. In this more abstract approach, the
spectral parameters of the Potts model are represented in terms of the
eigenvalues of the central Hopf subalgebra of  $U_q(\hat{s\ell} (2))$ ($q=
e^{2\pi i /N}$) with the algebraic curve fixed by the intertwining condition.

In a recent letter \refto{6}, we have discovered a new solution to the
Yang--Baxter equation, with the spectral parameters living on an algebraic
curve. This solution was obtained as the intertwiner for semi-cyclic
representations of the Hopf algebra  $U_q({s\ell} (2))$ for $q$ a third root of
unit and it shares some of the generic properties of factorizable
$S$--matrices.
In this paper, we generalize the results of \refto{6} for $q^N=1$, $N\ge5$. The
case $N=4$ is special and is studied separately in \refto{12}. Inspired by the
chiral Potts model, we propose a solvable lattice model whose Boltzmann weights
are identified with the  $U_q({s\ell} (2))$ intertwiners for semi-cyclic
representations. Since we are working with  $U_q({s\ell} (2))$
and not its affine extension, and based on the very simple structure of the
spectral manifold, we conjecture that the models we describe correspond to
critical points.

The organization of the paper is as follows. In section 2 we review the chiral
Potts model from the point of view of quantum groups. In section 3 we give the
intertwiner for semi-cyclic representations of  $U_q({s\ell} (2))$ for $N\ge3$
($q^N=1$). In section 4 we construct a solvable lattice model whose Boltzmann
weights are given by the previous intertwiner. In section 5 we study the
decomposition rules of tensor products of semi-cyclic representations, and
finally in section 6 we consider the $N\to\infty$ limit of the intertwiners of
semi-cyclic representations which leads to an infinite--dimensional
representation of the braid group.  This representation also satisfies the
Turaev condition for defining an invariant of knots and links.

\subhead{2. Chiral Potts as a model for affine  $U_q(\hat{s\ell} (2))$
intertwiners}

To define the $N$--chiral Potts model we assign to the sites of a square
lattice two different kinds of state variables: a $Z_N$ variable $m$
($=0,1,\ldots,N-1$) and a neutral variable $*$. On these variables one defines
the action of the group $Z_N$ as $\sigma(m)=m+1$ and $\sigma(*)=*$. The
allowed configurations for two adjacent sites of the lattice are of the kind
$(*,m)$,  \ie with one of the site variables the $Z_N$ neutral element $*$.
The rapidities $p,q,\ldots$ are associated with each line of the dual lattice
(figure~1).

\Fig 4 5.5 fig1 600 {The chiral Potts model: $p_i$ and $q_i$ are
rapidities, whereas $m,n,\ldots\in Z_N$}

 Correspondingly, there are two types of Boltzmann weights
represented graphically as
$$\eqalign{\ \cr W_{pq} (m-n) = \qquad\qquad\qquad \qquad\cr \ \cr \bar{W}_{pq}
(m-n) = \qquad \qquad\qquad\qquad\cr}  \eqno(1)$$

\vskip.5cm
\hskip8cm
\special{picture fx scaled 300}

The star--triangle relation of the model is
$$\sum_{d=1}^N {\bar W} _{qr} (b-d) W_{pr} (a-d) {\bar W}_{pq} (d-c) =
R_{pqr} W_{pq}(a-b) {\bar W}_{pr}(b-c) W_{qr} (a-c) \tag $$

Representing the vector rapidity by $(a_p,b_p,c_p,d_p)\in C$, it is found that
the star triangle relations of the model restrict the rapidities to lie on the
intersection of two Fermat curves: $$\eqalign{ a_p^N + k'b_p^N = k d_p^N \cr k'
a_p^N + b_p^N = k c_p^N \cr k^2 + {k'}^2 =1 \cr} \eqno(3)$$

In reference \refto{8} the previous model has been interpreted as describing a
scattering of kinks defined as follows. To each link of the lattice one
associates a kink operator which can be of the type $K_{*,n}(p)$ or
$K_{n,*}(p)$. This kink operators can be interpreted, in the continuum, as
configurations which interpolate between two extremal of some potential {\sl
\`a
la} Landau--Ginzburg, and which move with a rapidity $p$.

The Boltzmann weights of the lattice model are used in this interpretation to
define the $S$--matrix for these kinks:
$$\eqalign{ S\ket{K_{n*}(q) K_{*m}(p) } = W_{pq}(m-n) \ket{K_{n*}(p) K_{*m}(q)
} \cr S\ket{K_{*m}(q) K_{m*}(p) } = {\bar W}_{pq}(m-n) \ket{K_{*n}(p) K_{n*}(q)
} \cr} \tag $$
These models can be characterized  very nicely in connection with the
quantum affine extension of ${s\ell(2)}$, namely $U_\epsilon (
\hat{{s\ell(2)}}) $ with
$\epsilon = e^{2\pi i /N}$. First of all, the finite--dimensional irreps of
 $U_\epsilon ( \hat{{s\ell(2)}}) $ are parametrized by the eigenvalues of the
central subalgebra which is generated, in addition to the Casimir, by
$E_i^{N'}$,   $F_i^{N'}$ and $K_i^{N'}$, where $i=0,1$ and $N'=N$ if $N$ is
odd, and $N'=N/2$ if $N$ is even. The representations where $x_i =
E_i^{N'}$,    $y_i = F_i^{N'}$ and $z_i =K_i^{N'}$ are all different from zero,
have dimension $N'$ and are called cyclic \refto{4} or periodic \refto{9}
representations. The intertwining condition for the tensor product of cyclic
representations force these parameters $x_i, y_i, z_i$ to lie on an algebraic
curve which factorizes into two copies of the curve \(3). The corresponding
intertwiner $R$--matrix admits a representation as the product of four
Boltzmann weights of the chiral Potts model, actually two of them are chiral
$(W)$ and the other two antichiral $(\bar{W}$)  (see figure 2 and reference
\refto{4} for details).

\Fig 4 3.5 fig2 600 {The $R$--matrix for $U_q(\hat{s\ell} (2))$ and its chiral
Potts interpretation as a product of four Boltzmann weights}

 The need of four Boltzmann weights can be intuitively understood comparing the
structure of indices of a generic $R$--matrix $R_{r_1 r_2} ^{r'_1 r'_2} (\xi_1
,
\xi_2)$ with the Boltzmann weights \(1).  In fact, using only four Boltzmann
weights and therefore two independent rapidities one has enough degrees of
freedom to match the indices of the affine $R$--matrix in terms of lattice
variables. This is the formal reason for using the affine Hopf algebra to
describe the chiral Potts model.

Returning to the kink interpretation, one is associating to each $N'$--irrep of
$U_\epsilon ( \hat{{s\ell(2)}}) $ a two--kink  state with two different
rapidities.

\subhead{3. Intertwiners for semi-cyclic irreps of $U_\epsilon ( {{s\ell(2)}})
$ with
$\epsilon = e^{2\pi i /N}$, $N\ge3$}

In reference \refto{6} the intertwiner for semi-cyclic representations in the
case $N=3$ was considered. We shall give now the result for $N\ge3$. The case
$N=4$, \ie $q^2=1$, has more structure and is analyzed separately \refto{12}.

Let us first fix notation, essentially as in \refto{5}. We consider the Hopf
algebra $U_\epsilon ( {{s\ell(2)}}) $  with $\epsilon = e^{2\pi i /N}$ (we use
the
letter $\epsilon$ instead of $q$ in order to distinguish the case of $q$ a root
of unit), generated by $E$, $F$ and $K$ subject to the relations
$$\eqalign{ & EF- \epsilon^2  FE = 1-K^2 \cr & KE= \epsilon^{-2} EK \cr & KF =
\epsilon^2 FK \cr} \tag 4$$
and co-multiplication
$$\eqalign{ &\Delta E= E\otimes 1 + K \otimes E \cr & \Delta F = F \otimes 1 +
K \otimes F \cr & \Delta K = K \otimes K \cr } \tag 5$$ Notice that we do not
include in the commutator between $E$ and $F$ the usual denominator
$1-\epsilon^{-2}$.

When $\epsilon = e^{2\pi i /N}$ the central Hopf subalgebra $Z_\epsilon$ is
generated by $x= E^{N'}$, $z=F^{N'}$ and $z=K^{N'}$ where $N'=N$ ($N$ odd) or
$N'=N/2$ ($N$ even). We shall be interested in the special class of
representations for which $x=0$ but $ y$ and $z=\lambda^{N'}\not= \pm 1$ are
arbitrary non--zero complex numbers. These are the so--called semi-cyclic or
semi-periodic representations \refto{6}. Denoting by $\xi$ the couple of
values $(y,\lambda)$ which characterizes a semi-cyclic representation, then the
problem is to find a matrix $R(\xi_1, \xi_2)$ which intertwines between the
tensor products $\xi_1 \otimes \xi_2$ and $\xi_2 \otimes \xi_1$. Let $V_\xi$ be
the representation space associated with the semi-cyclic representation $\xi$,
which is spanned by a basis $\{ e_r(\xi)\}_{r=0}^{N'-1}$, then the intertwiner
$R$--matrix is an operator $R:V_{\xi_1} \otimes V_{\xi_2} \to V_{\xi_2} \otimes
V_{\xi_1}$ :
$$R(\xi_1, \xi_2) e_{r_1} (\xi_1) \otimes e_{r_2} (\xi_2) = R_{r_1 r_2 }^{r'_1
r'_2}  e_{r'_1} (\xi_2) \otimes e_{r'_2} (\xi_1) \tag $$ which satisfies the
equation
$$ R(\xi_1,\xi_2)  \Delta_{\xi_1\xi_2}(g) = \Delta_{\xi_2\xi_1}(g)
R(\xi_1,\xi_2)  \qquad [\forall g\in U_\epsilon({s\ell(2)}] \eqno(7)$$
where $ \Delta_{\xi_1\xi_2}(g)$ reflects the action of the quantum operator $g$
on $V_{\xi_1}\otimes V_{\xi_2}$:
$$ \Delta_{\xi_1\xi_2}(g) \left( e_{r_1} (\xi_1) \otimes  e_{r_2} (\xi_2)
\right)   = \Delta_{\xi_1\xi_2}(g) _{r_1r_2}^{r'_1r'_2} e_{r'_1} (\xi_1)
\otimes  e_{r'_2} (\xi_2)  \tag$$
Equation \(7) then reads explicitely as $$R_{r'_1r'_2} ^{s_1  s_2}
(\xi_1,\xi_2)  \Delta_{\xi_1\xi_2}(g) _{r_1r_2}^{r'_1r'_2} =
\Delta_{\xi_2\xi_1}(g) _{r'_1r'_2} ^{s_1  s_2} R  _{r_1r_2}^ {r'_1r'_2}
(\xi_1,\xi_2) \tag $$ Hence our convention is that contracted indices are
summed up in the SW--NE direction. An alternative way to write eq. \(7) is in
terms of the matrix ${\cal R}(\xi_1,\xi_2)  = P  R(\xi_1,\xi_2) $ where  $P:
V_{\xi_1} \otimes V_{\xi_2} \to V_{\xi_2} \otimes V_{\xi_1}$ is the permutation
map. Note that ${\cal R}  _{r_1r_2}^{r'_1r'_2} =   R _{r_1r_2}^{r'_2r'_1} $.
Equation \(7) reads then $${\cal R}(\xi_1,\xi_2)  \Delta_{\xi_1\xi_2}(g) =
\left( \sigma \circ  \Delta_{\xi_1\xi_2}(g)\right) {\cal R}(\xi_1,\xi_2) \tag
10
$$ where $\sigma $ is the permutation map $\sigma(a\otimes b) = b\otimes a$ of
the Hopf algebra. The universal form of \(10) is in fact one of the defining
relations of a quantum group, but as will be clear soon, we shall be working at
the representation level, without assuming the existence of a universal $\cal
R$--matrix.

After these preliminaries, the first important result that one derives from
\(7), when applied to an element $g$ belonging to the center $Z_\epsilon$ of
the algebra, is the following constraint on the possible values of $\xi_1$ and
$\xi_2$:
$${y_1  \over 1-\lambda_1^{N'}} = {y_2  \over 1-\lambda_2^{N'}} =k \eqno(11)$$
with $k$ an arbitrary complex number different from zero. When $N$ is odd, an
explicit form of a semi-cyclic representation satisfying $y=k(1-\lambda^N)$ is
given in the basis $\{e_r\}_{r=0}^{N-1}$ by\footnote{$^*$}{In reference
\refto{6}
 we used another basis $\tilde{e}_r$ which is related to the one we use now by
$$ {\tilde e}_r = k^{r/N}  \prod_{\ell=0}^{r-1} (1-\lambda \epsilon^\ell) e_r
$$ where the generator $F$ acts as $F{\tilde e}_r = {\tilde e}_{r+1}$ for $0\le
r\le N-2$ and $F{\tilde e}_{N-1}=k {\tilde e}_0$. In going to the new basis
$e_r$, we want to exploit the cyclicity of the generator $F$: in fact, we may
identify $e_r$ with $e_{r+N}$.}
$$\eqalign{&Fe_r = k^{1/N} (1-\lambda \epsilon^r) e_{r+1}  \cr
&Ee_r = {1\over k^{1/N}} [r] (1+\lambda \epsilon^{r-1}) e_{r-1}  \cr
&Ke_r = \lambda \epsilon^{2r} e_r \cr} \eqno(12)$$
where $[r]$ is the $q$--number
$$[r]= {1-\epsilon ^{2r} \over 1-\epsilon^2} \tag $$
Notice that this representation is highest weight, with $e_0$ the highest
weight vector.

It will be convenient for the rest of our computations to introduce the
following numbers
$$(\lambda)_{r_1,r_2} = \cases { \prod _{\ell=r_1}^{r_2-1} (1-\lambda
\epsilon^\ell) &  if $r_1<r_2$ \cr &\cr {[r_1]! \over [r_2]! }
\prod _{\ell=r_2}^{r_1-1} (1+\lambda
\epsilon^\ell) &  if $r_1>r_2$ \cr &\cr 1& if $r_1=r_2$ \cr} \eqno()$$
in terms of which
$$\eqalign{ F^n e_r = (\lambda)_{r,r+n}\ e_{r+n} \cr
E^n e_r = (\lambda)_{r,r-n}\ e_{r-n} \cr}\eqno()$$ They satisfy the obvious
relation $$ (\lambda)_{r_1,r_2} (\lambda)_{r_2,r_3} = (\lambda)_{r_1,r_3}
\qquad \matrix{ r_1\ge r_2\ge r_3 \cr r_1\le r_2\le r_3 \cr }\eqno()$$ and the
less obvious one which reflects the quantum algebra \(4):
$$\eqalign{  (\lambda)_{r,r+1} (\lambda)_{r+1,n}  - \epsilon^{2(r+1-n)}
(\lambda)_{r,n} (\lambda)_{n,n-1} (\lambda)_{n-1,n}\cr  = {[r-n+1] \over
[r+n+1]
} (\lambda)_{r,n} (\lambda)_{r+n+1,r+n} (\lambda)_{r+n,r+n+1}\cr} \quad (0\le
n\le r)\eqno()$$  Using now the representation \(12) we shall look for
intertwiners $R(\xi_1,\xi_2)$ satisfying the Yang--Baxter equation
$$ \left( {\bf 1} \otimes R(\xi_1,\xi_2) \right)
\left( R(\xi_1,\xi_3)   \otimes{\bf 1}\right)
 \left( {\bf 1} \otimes R(\xi_2,\xi_3) \right)  =
\left( R(\xi_2,\xi_3)   \otimes{\bf 1}\right)
 \left( {\bf 1} \otimes R(\xi_1,\xi_3) \right)
\left( R(\xi_1,\xi_2)   \otimes{\bf 1}\right) \eqno(18)$$ which guarantees a
unique intertwiner between the representations $\xi_1 \otimes \xi_2 \otimes
\xi_3$ and $\xi_3 \otimes \xi_2 \otimes \xi_1$.

In reference \refto{6}, it was found (when $N=3$) that an intertwiner $R$
matrix satisfying the Yang--Baxter equation \(18) exists and is unique, and
that in addition it satisfies the following three properties:

\item{i} Normalization.
$$R(\xi,\xi)={\bf 1} \otimes {\bf 1} \tag 19$$

\item{ii} Unitarity.
$$R(\xi_1,\xi_2) R(\xi_2,\xi_1) = {\bf 1} \otimes {\bf 1} \tag 20$$

\item{iii} Reflection symmetry
$$R(\xi_1,\xi_2) =P R(\xi_2,\xi_1) P  \tag 21$$

Among these properties, the last one is the most important and, as we shall
see, it is the key to find solutions for $N\ge 3$ (the case $N=4$, thus $N'=2$,
is singular in this crucial point, see \refto{12}). Explicitly, condition \(21)
reads
$$R_{r_1r_2}^{r'_1r'_2} (\xi_1,\xi_2)  =
R_{r_2r_1}^{r'_2r'_1} (\xi_2,\xi_1)  \tag $$

The strategy for finding solutions to the Yang--Baxter equation \(18) is to
look for intertwiners satisfying \(7) and having the reflection symmetry \(21).
We also normalize the $R$ matrix to be one when acting on the vector
$e_0\otimes  e_0$ (\ie, $R_{00}^{00}(\xi_1,\xi_2)=1$):
$$ R(\xi_1, \xi_2) e_0 (\xi_1) \otimes e_0 (\xi_2)  =  e_0(\xi_2) \otimes e_0
(\xi_1) = P e_0 (\xi_1) \otimes e_0 (\xi_2) \eqno(22)$$
The procedure we follow consists of finding, from the intertwiner condition
\(7) and the reflection symmetry \(21), a set of recursive equations for the
$R$--matrix which can be finally solved using the equation \(22). Indeed,
introducing \(21) into \(7), we obtain
$$ R(\xi_1, \xi_2) P \Delta _{\xi_2\xi_1}(g) P = P \Delta _{\xi_1\xi_2}(g) P
R(\xi_1, \xi_2) \tag 23$$
Specializing \(7) and \(23) to the case $g=F$, we obtain a set of equations
which can be solved to yield the following recursion formulae:
$$\eqalign{ R(\xi_1, \xi_2) (F_1\otimes 1 ) = \left[ A R(\xi_1, \xi_2)  - B
R(\xi_1, \xi_2)  (K_1 \otimes 1)  \right] {1\over 1-K_1 \otimes K_2}  \cr
R(\xi_1, \xi_2) (1\otimes F_2 ) = \left[ B R(\xi_1, \xi_2)  - A R(\xi_1, \xi_2)
(1\otimes K_2)  \right] {1\over 1-K_1 \otimes K_2}  \cr
}\tag 24$$ where $A$ and $B$ are two commuting operators given by
$$\eqalign{ & A= \Delta_{\xi_2,\xi_1} (F) = F_2 \otimes 1 + K_2 \otimes F_1 \cr
 & B= (\sigma \circ \Delta)_{\xi_2,\xi_1} (F) = F_2 \otimes K_1 + 1 \otimes F_1
\cr }\tag $$
The subindices in $F_1$, $F_2$, etc. are in fact unnecessary if we recall that
the whole equation is defined acting on the space $V_{\xi_1} \otimes
V_{\xi_2}$.

Iterating eqs. \(24) we find $$\eqalign{ &R(\xi_1, \xi_2)  (F_1^{r_1} \otimes
F_2^{r_2} )  =  \cr& \left( \sum_{s_1=0}^{r_1}  (-1)^{s_1} \epsilon
^{s_1(s_1-1)}  \left[\matrix{r_1 \cr s_1\cr}\right]  A^{r_1-s_1} B^{s_1}
\right) \left( \sum_{s_2=0}^{r_2}  (-1)^{s_2} \epsilon
^{s_2(s_2-1)}  \left[\matrix{r_2 \cr s_2\cr}\right]  B^{r_2-s_2} A^{s_2}
\right)   \cr  &\qquad\qquad \times R(\xi_1, \xi_2) (K_1^{s_1} \otimes
K_2^{s_2}
) {1\over \prod_{\ell=0}^{r_1+r_2-1} (1-\epsilon^{2\ell} K_1 \otimes K_2) }
 \cr} \tag 26$$
The final result is obtained by applying the above equation to the vector
$e_0(\xi_1) \otimes  e_0(\xi_2)$ and making use of \(22):
$$\eqalign{ &R(\xi_1, \xi_2)  (F_1^{r_1} e_0(\xi_1) \otimes
F_2^{r_2} e_0(\xi_2))  =  \cr& {1\over
\prod_{\ell=0}^{r_1+r_2-1} (1-\epsilon^{2\ell} \lambda_1 \lambda_2) }
 \prod_{\ell_1=0}^{r_1-1}  (A-\lambda_1 \epsilon^{2\ell_1} B)
\prod_{\ell_2=0}^{r_2-1}  (B-\lambda_2 \epsilon^{2\ell_2} A )
e_0(\xi_2) \otimes  e_0(\xi_1)\cr} \eqno(27)$$ In deriving \(27), we have used
the Gauss binomial formula
$$\sum_{\nu=0}^n (-1)^\nu \left[\matrix{n \cr \nu\cr}\right] z^\nu
\epsilon^{\nu(\nu-1)} = \prod_{\nu=0}^{n-1} (1-z\epsilon^{2\nu}) \tag $$

Letting $r_1$ or $r_2$ be $N'$ in \(27), one derives the intertwining
conditions \(11) which means that the spectral manifold for the solution \(27)
is fiven by the genus zero algebraic curve $$y=k(1-\lambda^{N'})  \tag 29$$

The general solution \(27) is valid for $N$ odd or even and it coincides, up to
a change of basis, with the one presented in \refto{6} for the particular case
$\epsilon^3=1$.

Expanding \(27) one can find the entries $R_{r_1r_2}^{r'_1r'_2}$ as functions
of the spectral variables $\xi_1$ and $\xi_2$. Some of them are easy to compute
since they are products of monomials, for example $(0\le n\le r)$
$$R_{r,0}^{r-n,n} (\xi_1, \xi_2) = { (\lambda_1)_{r,n} (\lambda_2) _{0,r-n}
\over [r-n]! } { \prod _{\ell=0}^{n-1} (\lambda_2  - \lambda_1
\epsilon^{2\ell})
\over \prod _{\ell=0}^{r-1} (1  - \lambda_1 \lambda_2 \epsilon^{2\ell}) }
\eqno(30)$$
but in general they have a polynomial structure as in
$$R_{11}^{11} (\xi_1, \xi_2) = 1 - [2] {(\lambda_1-\lambda_2)^2 \over
(1-\lambda_1 \lambda_2) (1-\epsilon^2 \lambda_2 \lambda_2) } \eqno()$$
We shall give a general formula of $R_{r_1r_2}^{r'_1r'_2}$ when we consider
the limit $N\to\infty$.

Another property of the $R$ matrix is the conservation of the quantum number
$r$ modulo $N'$, \ie
$$ R_{r_1r_2}^{r'_1r'_2} =0 \quad {\rm unless} \quad r_1+r_2 = r'_1 +r'_2
\ {\rm mod }\ N' \eqno()$$ The previous construction reveals that the
intertwining condition, when supplemented with the reflection relation \(21),
are enough data to produce solutions to the Yang--Baxter equation. The general
solution also satisfies the normalization \(19) and unitarity conditions \(20).

\subhead{4. A solvable lattice model for $U_\epsilon({s\ell(2)})$ intertwiners}
Our next task will be to define a lattice model whose Boltzmann weights admit a
direct interpretation in terms of intertwiners for semi-cyclic representations.
For these models, the $R$--matrix given in \(27) is the solution to the
associated star--triangle relation. To define the model we associate to each
site of the lattice a $Z_N$ variable $m$ such that $\sigma (m) = m+1$ with
$\sigma$ the generator of $Z_N$ transformations.  In the same way as for chiral
Potts, we associate with each line of the dual lattice a rapidity which we
denote by $\xi$. The Boltzmann weight for the plaquette is then defined by

$$W_{\xi_1\xi_2} (m_1,m_2,m_3,m_4) = R_{r_1r_2}^{r'_1r'_2} (\xi_1,\xi_2) =
\qquad \eqno(33)$$

\hskip11cm
\special{picture figx2 scaled 300}

$$ r_1=m_1-m_2,  \quad r_2= m_2-m_3,  \quad r'_1  = m_1-m_4, \quad r'_2 =
m_4-m_3
$$

The $R$--matrix in \(33) is the intertwiner for two semi-cyclic representations
$ \xi_1$ , $\xi_2$. The rapidities are now two--vectors $\xi=(y,\lambda)$ and
they are forced to live on the algebraic curve \(29). The definition \(33) is
manifestly $Z_N$ invariant: $m_i\to \sigma (m_i)$. From \(33) it is also easy
to see that the star--triangle relation for the $W$'s becomes the Yang--Baxter
equation for the $R$'s.

The equivalent here to the kink interpretation can be easily done substituting
the kinks by some solitonic--like structure. In fact, if we interpret the
latice variables $m$ as labelling different vacua connected  by $Z_N$
transformations, we can interpret in the continuum each link as representing an
interpolating configuration between two different vacua.  $e_r$ is identified
with a soliton configuration connecting the two vacua $m$ and $n$ with
$r=m-n$. In these conditions the Boltzmann weights \(33) become the scattering
$S$--matrix for two of these configurations with rapidities $\xi_1$ and $\xi_2$
(see fig.~3).

\Fig 4 2 fig3 600 {$S$--matrix interpretation of the $U_q({s\ell} (2))$
$R$--matrix \(27)}

 Notice that the conditions we have used in the previous section
for solving the Yang--Baxter equation are very natural in this solitonic
interpretation. In fact, the reflection relations \(21) simply mean that the
scattering is invariant under parity transformations. The condition
$R_{00}^{00}=1$ can now be interpreted as some kind of vacuum stability. Notice
that this condition is very dependent on the highest weight vector nature of
semi-cyclic representations. Moreover, the normalization condition \(19) and
the
unitarity property \(20) strongly support the $S$--matrix interpretation.  The
main difference between the model \(33) and chiral Potts is that for \(33)
we allow for all sites in the same plaquette arbitrary $Z_N$ variables. In this
way we loose the chiral difference between vertical and horizontal interactions
($W,\bar W$). What we gain is the possibility to connect the
$U_\epsilon({s\ell(2)})$ intertwiners with Boltzmann weights without passing
through the affine extension.

\subhead{5. Decomposition rules for semi-cyclic representations: the bootstrap
property}

One of the main ingredients used in ref. \refto{4} to obtain solutions of the
Yang--Baxter equation was the fact that generic cyclic representations of
$U_\epsilon(\hat{{s\ell(2)}})$ are indecomposable. This is not the case for
semi-cyclic representations of $U_\epsilon({s\ell(2)})$ \refto{6,9}. The
decomposition rules are given by
$$(\lambda_1,y_1)\otimes (\lambda_2, y_2) = \bigoplus_{\ell=0}^{N'-1}
(\epsilon^{2\ell} \lambda_1 \lambda_2 , y_1+ \lambda_1^{N'} y_2)   \eqno(34)$$
If we consider the tensor product in the reverse order,
$$(\lambda_2,y_2)\otimes (\lambda_1, y_1) = \bigoplus_{\ell=0}^{N'-1}
(\epsilon^{2\ell} \lambda_1 \lambda_2 , y_2+ \lambda_2^{N'} y_1)   \eqno()$$
we deduce that the irreps appearing in  $\xi_1\otimes \xi_2$ and
 $\xi_2\otimes \xi_1$ are the same provided the spectral condition \(11) is
satisfied. This is of course related to the existence of an intertwiner between
the two tensor products. We also observe that the tensor product of irreps on
the algebraic variety \(29) decomposes into irreps belonging to the same
variety. To obtain these rules, we simply need to use the co-multiplication
laws \(5) and the relation $$\Delta_{\xi_1 \xi_2} (g) K_{\xi_1\xi_2}^\xi =
 K_{\xi_1\xi_2}^\xi \rho_\xi(g) \qquad \forall g \in U_\epsilon({s\ell(2)})
\eqno()$$ with $ K_{\xi_1\xi_2}^\xi$ the Clebsch--Gordan projector
$ K_{\xi_1\xi_2}^\xi: V_\xi \to V_{\xi_1} \otimes V_{\xi_2} $ which exists
whenever $\xi\subset \xi_1 \otimes \xi_2$. Writing now $$e_r(\xi) =
K_{\xi_1\xi_2r}^{r_1r_2\xi} e_{r_1} (\xi_1) \otimes  e_{r_2} (\xi_2)
\tag $$ we obtain for the CG coefficients
$K_{\xi_1\xi_20}^{r_1r_2\xi^{(\ell)}}$, which give the highest weight vector
$e_0(\xi)$ of the irrep $\xi^{(\ell)} = ( \epsilon^{2\ell} \lambda_1 \lambda_2,
y_{12})$ in the semi-cyclic basis \(12), the following expression with
$\ell=r_1+r_2$: $$\eqalign{ K_{\xi_1\xi_20}^{r_1r_2\xi^{(\ell)}} &= (-1)^{r_1}
\epsilon^{r_1(r_1-1)}  \left[ \matrix{ \ell\cr r_1\cr} \right] \lambda_1^{r_1}
\prod_{r_1}^{\ell-1} (1+\lambda_1 \epsilon^\nu) \prod_{r_2}^{\ell-1} (1 -
\lambda_2 \epsilon^\nu) \cr &= {(-1)^{r_1}\over [\ell]!}  \epsilon^{r_1(r_1-1)}
\lambda_1^{r_1}\ (\lambda_1)_{\ell,r_1}\ (\lambda_2)_{\ell,r_2} \cr}
\eqno(38)$$
Coming back to the $S$--matrix picture the decomposition rules \(34) should be
interpreted as reflecting some kind of bootstrap property. This naive
interpretation is not quite correct. In fact, the decomposition rule \(34)
implies a strictly quantum composition of the rapidities determined by the Hopf
algebra structure of the center $Z_\epsilon$ of $U_\epsilon({s\ell(2)})$ at
$\epsilon$ a root of unit. This is a new phenomenon  derived from the fact
that we are considering something that looks like a factorized $S$--matrix for
particles, but where the kinematical properties of these particles, the
rapidities, are quantum group eigenvalues and therefore their composition rules
are not classical. Moreover, from the explicit expression of the
Clebsch--Gordan \(38) we observe another interesting phenomenon of mixing
between what we should consider internal quantum numbers, those labelling the
basis for the representation (\ie, the $r$'s) with the kinematical ones, \ie
the ones labelling the irreps (see fig. 4). If this physical picture is correct
this is the first case where the quantum group appears not only at the level of
internal symmetries but also determines the kinematics.

\Fig 4 2.8 fig4 600 {Decomposition rules derived from the Clebsch--Gordan
coefficients \(38) with the quantum group decomposition of rapidities}

{}From a strictly quantum group point of view we can, using the intertwiner
solution \(27) and the CG coefficients \(38), check some of the standard
results in representation theory of quantum groups. As an interesting example
we consider the relation $$R(\xi_1,\xi_2) K_\xi ^{\xi_1\xi_2} =
\phi(\xi_1,\xi_2,\xi)  K_\xi ^{\xi_2\xi_1} \tag 39$$ which is true for regular
representations of spin $j$ with $\phi(j_1,j_2,j) = (-1)^{j_1+j_2-j}
\epsilon^{C_j-C_{j_1} -C_{j_2}}$ and $C_j = j (j+1)$ the classical Casimir. For
semi-cyclic representations and the case $N=3$ we have found that the factors
$\phi(\xi_1,\xi_2,\xi)$ in \(39) are equal to one, and we presume that this
fact persists for all $N$.

\subhead{6. The limit $N\to\infty$}

In this section we shall study the limit $N\to\infty$ of the $R$--matrix found
in section 3. The quantum deformation parameter $\epsilon$ goes in this limit
to~1, so one could expect that the Hopf algebra  $U_\epsilon ( {{s\ell(2)}}) $
becomes the classical universal enveloping algebra of ${s\ell(2)}$. This is not
however what is happening here as can be seen by taking $\epsilon\to1$ in eqs.
\(4):
$$[E,F]=1-K^2 \qquad [K,E]=[K,F]=0 \eqno(40)$$
Let us recall that in the quantum group relations  \(4) we have not included
the denominator $1-\epsilon^{-2}$, hence there is no need to apply the
L'H\^opital rule. What we obtain rather in the limit
$N\to\infty$ is the Heisenberg algebra of a harmonic oscillator. Inded,
defining
$a$ and $a^\dagger$ as
$$a={1\over 1+K} E  \qquad a^\dagger  = {1\over 1-K} F  \eqno()$$ we find that
\(40) amount to
$$[a,a^\dagger]=1 \qquad [K,a]=[K,a^\dagger]=0\eqno(42)$$ The role of the
operator $K$ is therefore to produce non--trivial co-multiplications preserving
the algebra \(40) or \(42), and this is why we may have non--trivial
$R$--matrices. The representation spaces of the algebra \(40) are now
infinite--dimensional and are labelled by the value of $K$; we shall call them
${\cal H}_\lambda$. In a basis $\{e_r\}_{r=0}^{\infty}$ of ${\cal H}_\lambda$
we have
$$\eqalign{ &Fe_r = (1-\lambda) e_{r+1}  \cr
&Ee_r =r (1+\lambda) e_{r-1}  \cr
&Ke_r = \lambda e_r \cr} \eqno(43)$$ Hence we see that $e_r$ can be identified
with the $r$--th level of a harmonic oscillator. Of couse the value $\lambda=1$
in \(43) has to be treated with care.

The $R$--matrix is now an operator $R(\lambda_1,\lambda_2)  : {\cal
H}_{\lambda_1}
\otimes {\cal H}_{\lambda_2} \to {\cal H}_{\lambda_2} \otimes {\cal
H}_{\lambda_1} $ which in the
basis \(43) has the following non--vanishing entries:
$$\eqalign{ &R_{r_1,r_2}^{r_1+r_2-\ell, \ell} (\lambda_1,\lambda_2)  =
{1\over (1-\lambda_1\lambda_2)^{r_1+r_2}}  \cr &\sum_{\ell_1+\ell_2=\ell} {r_1
\choose \ell_1} {r_2\choose \ell_2} [(1+\lambda_1)  (1-\lambda_2)]^{r_1-\ell_1}
 [(1-\lambda_1)  (1+\lambda_2)]^{\ell_2}  (\lambda_2-\lambda_1)^{\ell_1}
(\lambda_1-\lambda_2)^{r_2-\ell_2}  \cr}\tag$$ where $0\le \ell \le r_1+r_2$.
This expression has been obtained from \(27) taking the limit $\epsilon\to1$
and using \(43).

A first observation is that $R(\lambda_1,\lambda_2)$ depends only on the
following harmonic ratio $$\eta_{12} = {z_{12} z_{34} \over z_{13} z_{24}} =
2{\lambda_1-\lambda_2 \over (1+\lambda_1) (1-\lambda_2)}   \tag $$
which corresponds to a sphere with 4 punctures at the points $z_1=\lambda_1$, $
z_2=\lambda_2$, $z_3=-1$ and $z_4=1$. In these new variables we have
$$R_{r_1,r_2}^{r_1+r_2-\ell ,\ell} (\eta_{12}) = {1\over
(1-\eta_{12}/2)^{r_1+r_2}} \sum_{\ell_1+\ell_2=\ell} (-1)^{\ell_1} {r_1 \choose
\ell_1} {r_2\choose \ell_2}  (\eta_{12}/2)^{r_2-\ell_2+\ell_1}
(1-\eta_{12})^{\ell_2} \tag $$ A reflection transformation $\eta_{12} \to
\eta_{21} =  {\eta_{12} \over \eta_{12}-1}$ corresponds to a M\"obius
transformation.

It is however more convenient to use another variable $u$ defined as
$$u_{12} = {\lambda_1 -\lambda_2 \over 1-\lambda_1\lambda_2} = {\eta_{12} \over
2-\eta_{12}} \tag $$ which changes sign under a reflection symmetry.  The
$R$--matrix reads in the $u$--variable as
$$R_{r_1,r_2}^{r_1+r_2-\ell, \ell} (u_{12}) =
(1+u_{12})^{r_1} u_{12}^{r_2}\sum_{\ell_1+\ell_2=\ell} (-1)^{\ell_1} {r_1
\choose \ell_1} {r_2\choose \ell_2}  \left({u_{12}  \over
1+u_{12}}\right)^{\ell_1} \left({1-u_{12}\over u_{12}}\right)^{\ell_2} \tag $$
Recalling the definition of the Jacobi polynomials $P_n^{(\alpha,\beta)}(x)$
($n=0,1,\ldots$):
$$P_n^{(\alpha,\beta)}(x)= {1\over 2^n} \sum_{m=0}^n {n+\alpha \choose  m}
{n+\beta \choose n-m} (x-1)^{n-m} (x+1)^m \tag $$ we see finally that $R(u)$
can be written in the form
$$R_{r_1,r_2}^{r_1+r_2-\ell ,\ell} (u) =
(1+u)^{r_1-\ell} u^{r_2-\ell} P_\ell ^{(r_2-\ell,r_1-\ell)} (1-2u^2)  \tag
50$$
As an application of this formula we may derive the classical limit of eq.
\(30) knowing that $$P_\ell^{(-\ell,r-\ell)} (x) = {(-1)^\ell \over 2^\ell}
{r\choose \ell} (1-x)^\ell \tag $$

The reflection symmetry of $R$
$$R_{r_1,r_2}^{r_1+r_2-\ell, \ell} (u) =
R_{r_2,r_1}^{\ell, r_1+r_2-\ell } (-u) \tag $$ implies the following identity
between Jacobi polynomials:
 $$P_{\alpha+\beta+n}^{(-\alpha,-\beta)}(x) = \left({x-1\over2}\right)^\alpha
\left({x+1 \over 2}\right)^\beta  P_n^{(\alpha,\beta)}(x) \tag $$ whenever
$n+\alpha$, $n+\beta$ and $n+\alpha+\beta$ are all non--negative integers. The
Yang--Baxter equation of $R(u)$ implies a cubic equation for the Jacobi
polynomials that we shall not write down.

The fact that the $R$--matrix can be written entirely as a function of a single
variable $u$ has some interesting consequences. First of all, we see from the
definition of $u_{12}$ that interpreting $\lambda_i$ as the velocity of the
$i$-th soliton, then $u_{ij}$ is nothing but the relative speed of the $i$-th
soliton with respect to the $j$-th soliton in a $1+1$ relativistic world.
Moreover, the YB equation \(18) when written in the $u$ variables has a
relativistic look:
$$ \left( {\bf 1} \otimes R(u) \right)
\left( R\left({u+v\over 1+uv}\right)   \otimes{\bf 1}\right)
 \left( {\bf 1} \otimes R(v) \right)  =
\left( R(v)   \otimes{\bf 1}\right)
 \left( {\bf 1} \otimes R\left({u+v\over 1+uv}\right) \right)
\left( R(u)   \otimes{\bf 1}\right) \eqno(54)$$
and in particular the scattering matrices $R(u)$ of these relativistic solitons
give us a representation of the braid group provided $$u=v={u+v\over 1+uv}\tag
$$  a situation which happens when $u=0,\pm1$. The case $u=0$ is trivial, since
$R(u=0)=1$, but the other two yield the following braiding matrices:
$$\eqalign{ R_{r_1r_2}^{r'_1r'_2}(+)  = \lim_{u\to1} R_{r_1r_2}^{r'_1r'_2}(u)
= \delta_{r_1+r_2,r'_1+r'_2}  (-1)^{r'_2} 2^{r_1-r'_2} {r_1 \choose r'_2}  \cr
R_{r_1r_2}^{r'_1r'_2}(-)  = \lim_{u\to-1} R_{r_1r_2}^{r'_1r'_2}(u)
= \delta_{r_1+r_2,r'_1+r'_2}  (-1)^{r'_1} 2^{r_2-r'_1} {r_2 \choose r'_1}
\cr}\tag 56 $$ which satisfy, in addition to the YB equation, the relation
$$R(+)R(-)=1 \tag $$ which is a consequence of the unitarity condition \(20).

This means that we have obtained an infinite--dimensional representation $\pi$
of the braid group $B_n$ given by
$$\eqalign{ \pi: &\ B_n \to End({\cal H}^{\otimes n})  \cr & \sigma_i^{\pm1}
\mapsto 1\otimes \cdots \otimes R_{i,i+1}(\pm) \otimes \cdots \otimes 1
\cr}\tag $$  where ${\cal H}$ is isomorphic to the Hilbert space of a harmonic
oscillator. In fact, using the universal  matrix ${\cal R}=PR\in End
({\cal H}^{\otimes 2})$ we can write eqs. \(56) as   $$\eqalign{ {\cal R}(+)=
(e^{i\pi {\cal N}} \otimes 1)  e^{2a\otimes a^\dagger} \cr {\cal R}(-)=
(1\otimes
e^{i\pi {\cal N}} )  e^{2 a^\dagger\otimes a} \cr }\tag 59$$ where ${\cal
N}=a^\dagger a$ is
the number operator and $$a^\dagger e_r = e_{r+1} \qquad ae_r  = r e_{r-1}
\tag
$$ The Yang--Baxter relation \(54) in the limit $u\to\pm1$ can be most easily
proved in terms of the YB solution for the $\cal R$ matrix which reads $${\cal
R}_{12} {\cal R}_{13} {\cal R}_{23}  = {\cal R}_{23} {\cal R}_{13} {\cal
R}_{12} \tag $$ where ${\cal R}_{12} = (e^{i\pi {\cal N}} \otimes 1 \otimes 1)
e^{2a\otimes a^\dagger \otimes 1}\ $ etc.

We have thus seen that the limit $N\to \infty$ of our construction is well
defined, and has some interesting structure. It describes essentially the
scattering of relativistic solitons whose spectrum is that of a harmonic
oscillator. Interestingly enough, in the limit of very deep scattering one
obtains an infinite--dimensional representation of the braid group. One may
wonder whether this representation provides us with new invariants of knots and
links, just as the usual finite--dimensional $R$ matrices  from quantum groups
do.

With this in mind, we shall propose a slight generalization of the $\cal R$
matrices \(59) given by
$$\eqalign{ {\cal R}(x,y;+)=  (x^{ {\cal N}} \otimes y^{-{\cal N}})
e^{(y-x) a\otimes a^\dagger} \cr {\cal R}(x,y;-)=
e^{(x-y) a^\dagger\otimes a} (y^{\cal N}\otimes x^{-{\cal N}} ) \cr }\tag 62$$
where $x$ and $y$ are two independent complex numbers. It is easy to see that
these new $\cal R$ matrices also satisfy the YB equation, yielding a
representation $\pi_{x,y}: B_n \to End ({\cal H}^{\otimes n})$  of the braid
group.
The previous case is recoverd with $x=-1$, $y=1$.
The braiding matrices that follow from \(62)  are
$$\eqalign{ R_{r_1r_2}^{r'_1r'_2} (x,y)  = \delta_{r_1+r_2,r'_1+r'_2}
{r_1 \choose r'_2} (y-x)^{r_1-r'_2} x^{r'_2} y^{-r'_1}
\cr { R^{-1}}_{r_1r_2}^{r'_1r'_2} (x,y)  = \delta_{r_1+r_2,r'_1+r'_2}
{r_2 \choose r'_1} (x-y)^{r_2- r'_1} x^{-r_2} y^{r_1} \cr} \tag 63$$

This braid group representation admits an extension \`a la Turaev \refto{13},
\ie there exists an isomorphism $\mu:{\cal H}\to {\cal H}$ satisfying the
following three
conditions:
$$\eqalign{ i)&\quad  (\mu_i \mu_j -\mu_k \mu_\ell )\ R_{ij}^{k\ell}
=0 \cr
 ii)&\quad \sum_j  R_{ij}^{kj} \ \mu_j = \delta _i^k\ ab \cr
 iii)&\quad \sum_j  {R^{-1}}_{ij}^{kj}\  \mu_j = \delta _i^k\ a^{-1}b \cr
}\tag $$ for some constants $a$ and $b$. For the $R$ matrix \(63), the Turaev
conditions hold if $$\mu=1 \qquad a=b^{-1}= \sqrt{(y/x)}  \tag $$ The invariant
of knots and links that one would get is thus $$T_{x,y}(\alpha) =
(x/y)^{\12[w(\alpha)  -n]} {\rm tr}\, \pi_{x,y}(\alpha)  \tag 66$$ where
$\alpha\in B_n$ and $w(\alpha)$ is the writhe of $\alpha$. The trace in \(66)
is defined on the $n$-th tensor product of Hilbert spaces, therefore to make
sense of $T_{x,y}(\alpha)$ one should regularize this trace without  losing the
invariance under the Markov moves. We leave the identification and proper
definition of the invariant \(66) for a future publication.

\subhead{7. Final Comments}

A possible framework where to study in more detail the physical meaning of our
results could be the one recently developed by Zamolodchikov \refto{10} in
connection with the analysis of integrable deformations of conformal field
theories. The moral we obtain from our analysis is that a unique mathematical
structure, the quantum group, can describe at the same time conformal field
theories \refto{11} and integrable models.  The dynamics that fixes what of the
two kinds of physical systems is described is the way the central subalgebra
$Z_\epsilon$, for $q$ a root of unit, is realized. In the conformal case, the
central subalgebra is realized trivially with vanishing eigenvalues which
correspond to the regular representations. The quantum group symmetry is
defined in this case by the Hopf algebra quotiented by its central subalgebra.
{}From the results in \refto{3,4} and the ones described here it seems that
when
the center is realized in a non--trivial way the system we describe is an
integrable model. The star--triangle solution \(27) for the lattice model we
have defined in section 4 has good chances of describing a self--dual point.
The heuristic reason for this is that the algebraic curve \(29) on which the
spectral parameters live is of genus zero.

We want to mention also the possible physical implications in
this context of the recent mathematical results of reference \refto{5}. In
fact, these authors have defined a quantum co-adjoint action on the space of
irreps. This action divides the finite--dimensional irreps into orbits each one
containing a semi-cyclic representation. The co-adjoint action does not
preserve
the spectral manifold \(29) but if we maintain the interpretation of the irrep
labels as rapidities then it acts on them, opening in this way the door to new
kinematics completely based on quantum group properties.

Finally, we summarize the non--trivial results we have obtained in the limit
$N\to\infty$. First of all, the intertwiner matrix for semi-cyclic
representations can be interpreted as describing the scattering of solitons in
a $1+1$ relativistic world. The Hilbert space of these solitons is isomorphic
to that of the harmonic oscillator. Moreover, in the limit when the solitons
become relativistic, the scattering matrices provide us with an
infinite--dimensional representation of the braid group. This representation
seems to have a Markov trace which would allow us to find an invariant of links
and knots.

{\bf Acknowledgements}. We would like to thank M. Ruiz--Altaba for
continuing discussions on all aspects of this work and for sharing with us his
insights and results.  This work is partially supported by the Fonds National
Suisse pour la Recherche Scientifique.

\references

\refis{1} H. Au Yang, B.M. McCoy, J.H.H. Perk, S. Tang and M.~Yan,
\pl,A123,219,1987&.

\refis{2}R.J. Baxter, J.H.H. Perk and H. Au--Yang,
\pl,A128,138,1988&.

\refis{3}V.V.~Bazhanov and Yu.G.~Stroganov, \journal{J. Stat.
Phys.},59,799,1990&; V.V.~Bazhanov, R.M.~Kashaev, V.V.~Mongazeev and
Yu.G.~Stroganov, \cmp,138,393,1991&;  V.V.~Bazhanov and R.M.~Kashaev,
\cmp,136,607,1991&.

\refis{4}E. Date, M. Jimbo, M. Miki and T. Miwa, \pl,A148,45,1990&;
RIMS preprints 706, 715, 729 (1990).

\refis{5}C. de Concini and V. Kac, {\sl Quantum
group representations at $q$ a root of unity}, Pisa
preprint (May 1990); C.~de~Concini, V.~Kac and C.~Procesi,   Pisa
preprint (1991).

\refis{6}C. G\'omez, M. Ruiz--Altaba and G. Sierra, Gen\`eve preprint UGVA--DPT
1991/05--725 (May 1991), to appear in {\sl Phys. Lett.}~{\bf B}.

\refis{8}D. Bernard and V. Pasquier, {\sl Exchange algebra and exotic
supersymmetry in the chiral Potts model}, Saclay preprint SPhT/89--204.

\refis{9}D. Arnaudon, Ecole Polytechnique preprint (1991).

\refis{10} A.B. Zamolodchikov, in {\sl Knizhnik Memorial Volume}, eds. L.~Brink
\etal, World Scientific (1990) Singapore.

\refis{11}L. Alvarez--Gaum\'e, C. G\'omez and G. Sierra,  \np,B330,1990,347&;
C. G\'omez and G. Sierra,  \np,B352,1991,791&; C.~Ram\'{\i}rez, H.~Ruegg and
M.~Ruiz~Altaba, {\sl Nucl. Phys.}~{\bf B}, to appear.

\refis{12}M.Ruiz--Altaba, Gen\`eve preprint
UGVA--DPT~1991/08--741 (August 1991).

\refis{13}V.G. Turaev, \journal{Invent. Math.},92,527,1988&.

\endreferences

\endit